\input harvmac.tex 
 \input epsf.tex
 \input amssym


\def\figin{\epsfcheck\figin}\def\figins{\epsfcheck\figins}
\def\epsfcheck{\ifx\epsfbox\UnDeFiNeD
\message{(NO epsf.tex, FIGURES WILL BE IGNORED)}
\gdef\figin##1{\vskip2in}\gdef\figins##1{\hskip.5in}
\else\message{(FIGURES WILL BE INCLUDED)}%
\gdef\figin##1{##1}\gdef\figins##1{##1}\fi}
\def\DefWarn#1{}
\def\figinsert{\goodbreak\midinsert}
\def\ifig#1#2#3{\DefWarn#1\xdef#1{fig.~\the\figno}
\writedef{#1\leftbracket fig.\noexpand~\the\figno} %
\figinsert\figin{\centerline{#3}}\medskip\centerline{\vbox{\baselineskip12pt
\advance\hsize by -1truein\noindent\footnotefont{\bf
Fig.~\the\figno:} #2}}
\bigskip\endinsert\global\advance\figno by1}

\def\unit{\relax{\rm 1\kern-.26em I}}
\def\nada{\relax{\rm 0\kern-.30em l}}
\def\tilde{\widetilde}


\def \pa {\partial}

\def\p{\partial}

\noblackbox
\def\IL{\relax{\rm I\kern-.18em L}}
\def\IH{\relax{\rm I\kern-.18em H}}
\def\IR{\relax{\rm I\kern-.18em R}}
\def\IC{\relax\hbox{$\inbar\kern-.3em{\rm C}$}}
\def\IZ{\relax\ifmmode\mathchoice
{\hbox{\cmss Z\kern-.4em Z}}{\hbox{\cmss Z\kern-.4em Z}} {\lower.9pt\hbox{\cmsss Z\kern-.4em Z}}
{\lower1.2pt\hbox{\cmsss Z\kern-.4em Z}}\else{\cmss Z\kern-.4em Z}\fi}


\def\zb {\bar{z}}

\font\manual=manfnt \def\dbend{\lower3.5pt\hbox{\manual\char127}}

\def\ip{${\cal I}^+$}

\def\ci{{\cal I}}
\lref\bms{}
\lref\Thorne{
K.~S.~Thorne,
Physical Review D 45.2 (1992): 520.
}
\lref\GreeneYA{
  B.~R.~Greene, A.~D.~Shapere, C.~Vafa and S.~T.~Yau,
 ``Stringy Cosmic Strings and Noncompact Calabi-Yau Manifolds,''
Nucl.\ Phys.\ B {\bf 337}, 1 (1990)..
}
\lref\StromingerJFA{
  A.~Strominger,
  ``On BMS Invariance of Gravitational Scattering,''
JHEP {\bf 1407}, 152 (2014).
[arXiv:1312.2229 [hep-th]].
}
\lref\deBoerVF{
  J.~de Boer and S.~N.~Solodukhin,
  ``A Holographic reduction of Minkowski space-time,''
Nucl.\ Phys.\ B {\bf 665}, 545 (2003).
[hep-th/0303006].
}
\lref\BanksVP{
  T.~Banks,
 ``A Critique of pure string theory: Heterodox opinions of diverse dimensions,''
[hep-th/0306074].
}
\lref\BieriADA{
  L.~Bieri and D.~Garfinkle,
  ``A perturbative and gauge invariant treatment of gravitational wave memory,''
Phys.\ Rev.\ D {\bf 89}, 084039 (2014).
[arXiv:1312.6871 [gr-qc]].
}
\lref\PravdaVH{
  V.~Pravda and A.~Pravdova,
``Boost rotation symmetric space-times: Review,''
Czech.\ J.\ Phys.\  {\bf 50}, 333 (2000).
[gr-qc/0003067].
}
\lref\TolishBKA{
  A.~Tolish and R.~M.~Wald,
  ``Retarded Fields of Null Particles and the Memory Effect,''
[arXiv:1401.5831 [gr-qc]].
}

\lref\TolishODA{
  A.~Tolish, L.~Bieri, D.~Garfinkle and R.~M.~Wald,
  ``Examination of a simple example of gravitational wave memory,''
Phys.\ Rev.\ D {\bf 90}, 044060 (2014).
[arXiv:1405.6396 [gr-qc]].
}

\lref\CardosoPA{
  V.~Cardoso, O.~J.~C.~Dias and J.~P.~S.~Lemos,
  ``Gravitational radiation in D-dimensional space-times,''
Phys.\ Rev.\ D {\bf 67}, 064026 (2003).
[hep-th/0212168].
}

\lref\CachazoFWA{
  F.~Cachazo and A.~Strominger,
  ``Evidence for a New Soft Graviton Theorem,''
[arXiv:1404.4091 [hep-th]].
}
\lref\KapecOPA{
  D.~Kapec, V.~Lysov, S.~Pasterski and A.~Strominger,
  ``Semiclassical Virasoro symmetry of the quantum gravity $ \mathcal{S}$-matrix,''
JHEP {\bf 1408}, 058 (2014).
[arXiv:1406.3312 [hep-th]].
}
\lref\CheungIUB{
  C.~Cheung, A.~de la Fuente and R.~Sundrum,
  ``4D Scattering Amplitudes and Asymptotic Symmetries from 2D CFT,''
[arXiv:1609.00732 [hep-th]].
}
\lref\KapecJLD{
  D.~Kapec, P.~Mitra, A.~M.~Raclariu and A.~Strominger,
  ``A 2D Stress Tensor for 4D Gravity,''
[arXiv:1609.00282 [hep-th]].
}

\lref\CampigliaYKA{
  M.~Campiglia and A.~Laddha,
  ``Asymptotic symmetries and subleading soft graviton theorem,''
Phys.\ Rev.\ D {\bf 90}, no. 12, 124028 (2014).
[arXiv:1408.2228 [hep-th]].
}

\lref\GarfinkleEQ{
  D.~Garfinkle and A.~Strominger,
  ``Semiclassical Wheeler wormhole production,''
Phys.\ Lett.\ B {\bf 256}, 146 (1991).}

\lref\GarfinkleXK{
  D.~Garfinkle, S.~B.~Giddings and A.~Strominger,
  ``Entropy in black hole pair production,''
Phys.\ Rev.\ D {\bf 49}, 958 (1994).
[gr-qc/9306023].}

\lref\KapecOPA{
  D.~Kapec, V.~Lysov, S.~Pasterski and A.~Strominger,
  ``Semiclassical Virasoro symmetry of the quantum gravity $ \cal{S}$-matrix,''
JHEP {\bf 1408}, 058 (2014).
[arXiv:1406.3312 [hep-th]].
}

\lref\BarnichLYG{
  G.~Barnich and C.~Troessaert,
  ``Finite BMS transformations,''
[arXiv:1601.04090 [gr-qc]].
}

\lref\EardleyAU{
  D.~M.~Eardley, G.~T.~Horowitz, D.~A.~Kastor and J.~H.~Traschen,
  ``Breaking cosmic strings without monopoles,''
Phys.\ Rev.\ Lett.\  {\bf 75}, 3390 (1995).
[gr-qc/9506041].
}

\lref\PodolskyAT{
  J.~Podolsky and J.~B.~Griffiths,
  ``Null limits of the C metric,''
Gen.\ Rel.\ Grav.\  {\bf 33}, 59 (2001).
[gr-qc/0006093].
}

\lref\GregoryHD{
  R.~Gregory and M.~Hindmarsh,
  ``Smooth metrics for snapping strings,''
Phys.\ Rev.\ D {\bf 52}, 5598 (1995).
[gr-qc/9506054].
}

\lref\penrose{
R.~Penrose, 
``The geometry of impulsive gravitational waves,"
General Relativity, Papers in Honour of J.~L.~Synge (1972): 101-115.
}

\lref\nutkupenrose{
Nutku, Y., and R. Penrose. "On impulsive gravitational waves." Twistor Newsletter 34.9-12 (1992): 4.
}

\lref\GleiserVT{
  R.~Gleiser and J.~Pullin,
  ``Are Cosmic Strings Gravitationally Stable Topological Defects?''
Class.\ Quant.\ Grav.\  {\bf 6}, L141 (1989)..
}

\lref\PodolskyCX{
  J.~Podolsky and J.~B.~Griffiths,
  ``The Collision and snapping of cosmic strings generating spherical impulsive gravitational waves,''
Class.\ Quant.\ Grav.\  {\bf 17}, 1401 (2000).
[gr-qc/0001049].
}

\lref\GriffithsHJ{
  J.~B.~Griffiths and P.~Docherty,
  ``A Disintegrating cosmic string,''
Class.\ Quant.\ Grav.\  {\bf 19}, L109 (2002).
[gr-qc/0204085].
}

\lref\AshtekarAR{
  A.~Ashtekar and T.~Dray,
  ``On the Existence of Solutions to Einstein's Equation With Nonzero Bondi News,''
Commun.\ Math.\ Phys.\  {\bf 79}, 581 (1981)..
}

\lref\PodolskyXH{
  J.~Podolsky and R.~Svarc,
  ``Refraction of geodesics by impulsive spherical gravitational waves in constant-curvature spacetimes with a cosmological constant,''
Phys.\ Rev.\ D {\bf 81}, 124035 (2010).
[arXiv:1005.4613 [gr-qc]].
}

\lref\StromingerJFA{
  A.~Strominger,
  ``On BMS Invariance of Gravitational Scattering,''
JHEP {\bf 1407}, 152 (2014).
[arXiv:1312.2229 [hep-th]].
}
\lref\HeZEA{
  T.~He, P.~Mitra and A.~Strominger,
  ``2D Kac-Moody Symmetry of 4D Yang-Mills Theory,''
[arXiv:1503.02663 [hep-th]].
}
\lref\CampigliaQKA{
  M.~Campiglia and A.~Laddha,
  ``Asymptotic symmetries of QED and WeinbergÕs soft photon theorem,''
JHEP {\bf 1507}, 115 (2015).
[arXiv:1505.05346 [hep-th]].
}
\lref\KapecENA{
  D.~Kapec, M.~Pate and A.~Strominger,
 ``New Symmetries of QED,''
[arXiv:1506.02906 [hep-th]].
}
\lref\StromingerPWA{
  A.~Strominger and A.~Zhiboedov,
  ``Gravitational Memory, BMS Supertranslations and Soft Theorems,''
JHEP {\bf 1601}, 086 (2016).
[arXiv:1411.5745 [hep-th]].
}

\lref\BarnichSE{
  G.~Barnich and C.~Troessaert,
  ``Symmetries of asymptotically flat 4 dimensional spacetimes at null infinity revisited,''
Phys.\ Rev.\ Lett.\  {\bf 105}, 111103 (2010).
[arXiv:0909.2617 [gr-qc]].
}

\lref\KapecOPA{
  D.~Kapec, V.~Lysov, S.~Pasterski and A.~Strominger,
  ``Semiclassical Virasoro symmetry of the quantum gravity $ \cal{S}$-matrix,''
JHEP {\bf 1408}, 058 (2014).
[arXiv:1406.3312 [hep-th]].
}

\lref\BarnichLYG{
  G.~Barnich and C.~Troessaert,
  ``Finite BMS transformations,''
[arXiv:1601.04090 [gr-qc]].
}

\lref\CompereJWB{
  G.~Compere and J.~Long,
  ``Vacua of the gravitational field,''
[arXiv:1601.04958 [hep-th]].
}

\lref\RobinsonZZ{
  I.~Robinson and A.~Trautman,
  ``Some spherical gravitational waves in general relativity,''
Proc.\ Roy.\ Soc.\ Lond.\ A {\bf 265}, 463 (1962)..
}

\lref\HoganXJ{
  P.~A.~Hogan,
  ``A Spherical impulse gravity wave,''
Phys.\ Rev.\ Lett.\  {\bf 70}, 117 (1993)..
}

\lref\penrosenutku{
Y.~Nutku and R.~Penrose,
 ``On impulsive gravitational waves," 
 Twistor Newsletter 34.9-12 (1992): 4.
}

\lref\VilenkinZS{
  A.~Vilenkin,
  ``Gravitational Field of Vacuum Domain Walls and Strings,''
Phys.\ Rev.\ D {\bf 23}, 852 (1981)..
}

\lref\BicakRB{
  J.~Bicak and B.~Schmidt,
  ``On the Asymptotic Structure of Axisymmetric Radiative Space-times,''
Class.\ Quant.\ Grav.\  {\bf 6}, 1547 (1989)..
}

\lref\Bicak{
J.~Bicak,  
``Is there a news function for an infinite cosmic string?" 
Astronomische Nachrichten 311.3 (1990): 189-192.
}

\lref\AchucarroNU{
  A.~Achucarro, R.~Gregory and K.~Kuijken,
  ``Abelian Higgs hair for black holes,''
Phys.\ Rev.\ D {\bf 52}, 5729 (1995).
[gr-qc/9505039].
}

\lref\HawkingZN{
  S.~W.~Hawking and S.~F.~Ross,
  ``Pair production of black holes on cosmic strings,''
Phys.\ Rev.\ Lett.\  {\bf 75}, 3382 (1995).
[gr-qc/9506020].
}

\lref\EmparanJE{
  R.~Emparan,
  ``Pair creation of black holes joined by cosmic strings,''
Phys.\ Rev.\ Lett.\  {\bf 75}, 3386 (1995).
[gr-qc/9506025].
}

\lref\MaldacenaXJA{
  J.~Maldacena and L.~Susskind,
  ``Cool horizons for entangled black holes,''
Fortsch.\ Phys.\  {\bf 61}, 781 (2013).
[arXiv:1306.0533 [hep-th]].
}

\lref\WittenFP{
  E.~Witten,
  ``Cosmic Superstrings,''
Phys.\ Lett.\ B {\bf 153}, 243 (1985)..
}

\lref\PolchinskiIA{
  J.~Polchinski,
  ``Introduction to cosmic F- and D-strings,''
[hep-th/0412244].
}

\lref\ChristensenWB{
  M.~Christensen, A.~L.~Larsen and Y.~Verbin,
  ``Complete classification of the string - like solutions of the gravitating Abelian Higgs model,''
Phys.\ Rev.\ D {\bf 60}, 125012 (1999).
[gr-qc/9904049].
}

\lref\HongGX{
  K.~Hong and E.~Teo,
  ``A New form of the C metric,''
Class.\ Quant.\ Grav.\  {\bf 20}, 3269 (2003).
[gr-qc/0305089].
}

\lref\HongDM{
  K.~Hong and E.~Teo,
  ``A New form of the rotating C-metric,''
Class.\ Quant.\ Grav.\  {\bf 22}, 109 (2005).
[gr-qc/0410002].
}

\lref\SladekWY{
  P.~Sladek and J.~D.~Finley, III,
  ``Asymptotic properties of the C-Metric,''
Class.\ Quant.\ Grav.\  {\bf 27}, 205020 (2010).
[arXiv:1003.1471 [gr-qc]].
}

\lref\BicakVZ{
  J.~Bicak and A.~Pravdova,
  ``Symmetries of asymptotically flat electrovacuum space-times and radiation,''
J.\ Math.\ Phys.\  {\bf 39}, 6011 (1998).
[gr-qc/9808068].
}

\lref\HongGX{
  K.~Hong and E.~Teo,
  ``A New form of the C metric,''
Class.\ Quant.\ Grav.\  {\bf 20}, 3269 (2003).
[gr-qc/0305089].
}

\lref\SladekWY{
  P.~Sladek and J.~D.~Finley, III,
  ``Asymptotic properties of the C-Metric,''
Class.\ Quant.\ Grav.\  {\bf 27}, 205020 (2010).
[arXiv:1003.1471 [gr-qc]].
}

\lref\BicakVZ{
  J.~Bicak and A.~Pravdova,
  ``Symmetries of asymptotically flat electrovacuum space-times and radiation,''
J.\ Math.\ Phys.\  {\bf 39}, 6011 (1998).
[gr-qc/9808068].
}

\lref\GarfinkleEQ{
  D.~Garfinkle and A.~Strominger,
  ``Semiclassical Wheeler wormhole production,''
Phys.\ Lett.\ B {\bf 256}, 146 (1991)..
}

\lref\Cornish{
F.~H.~J.~Cornish, B.~Micklewright,
``The news function for Robinson-Trautman radiating metrics,''
Classical and Quantum Gravity, (1999) 16(2), p.611.
}

\lref\tafel{
J.~Tafel, 
``Bondi mass in terms of the Penrose conformal factor," 
Classical and Quantum Gravity 17.21 (2000): 4397.
}

\lref\ChruscielCJ{
  P.~Chrusciel,
  ``Semiglobal existence and convergence of solutions of the Robinson-Trautman (two-dimensional Calabi) equation,''
Commun.\ Math.\ Phys.\  {\bf 137}, 289 (1991).
}

\lref\PodolskyDU{
  J.~Podolsky and M.~Ortaggio,
  ``Robinson-Trautman spacetimes in higher dimensions,''
Class.\ Quant.\ Grav.\  {\bf 23}, 5785 (2006).
[gr-qc/0605136].
}

\lref\gary{G.W. Gibbons, Quantized flux tubes in Einstein-Maxwell theory and noncompact internal spaces, in Fields and geometry, proceedings of 22nd Karpacz Winter School of Theoretical Physics: Fields and Geometry, Karpacz, Poland, Feb 17 - Mar 1, 1986, ed. A. Jadczyk (World Scientific, 1986).}

\lref\vonderGonnaSH{
  U.~von der Gonna and D.~Kramer,
  ``Pure and gravitational radiation,''
Class.\ Quant.\ Grav.\  {\bf 15}, 215 (1998).
[gr-qc/9711001].
}

\lref\GregoryHD{
  R.~Gregory and M.~Hindmarsh,
  ``Smooth metrics for snapping strings,''
Phys.\ Rev.\ D {\bf 52}, 5598 (1995).
[gr-qc/9506054].
}

\lref\HawkingZN{
  S.~W.~Hawking and S.~F.~Ross,
  ``Pair production of black holes on cosmic strings,''
Phys.\ Rev.\ Lett.\  {\bf 75}, 3382 (1995).
[gr-qc/9506020].
}

\lref\bms{H.~Bondi, M.~G.~J.~van der Burg and A.~W.~K.~Metzner,
  ``Gravitational waves in general relativity. 7. Waves from axisymmetric isolated systems,''
Proc.\ Roy.\ Soc.\ Lond.\ A {\bf 269}, 21 (1962);
  R.~K.~Sachs,
  ``Gravitational waves in general relativity. 8. Waves in asymptotically flat space-times,''
Proc.\ Roy.\ Soc.\ Lond.\ A {\bf 270}, 103 (1962)..
}

\lref\GarfinkleXK{
  D.~Garfinkle, S.~B.~Giddings and A.~Strominger,
  ``Entropy in black hole pair production,''
Phys.\ Rev.\ D {\bf 49}, 958 (1994).
[gr-qc/9306023].
}

\lref\PodolskyNN{
  J.~Podolsky,
  ``Exact impulsive gravitational waves in space-times of constant curvature,''
[gr-qc/0201029].
}

\lref\PodolskyCX{
  J.~Podolsky and J.~B.~Griffiths,
  ``The Collision and snapping of cosmic strings generating spherical impulsive gravitational waves,''
Class.\ Quant.\ Grav.\  {\bf 17}, 1401 (2000).
[gr-qc/0001049].
}

\lref\PodolskyXH{
  J.~Podolsky and R.~Svarc,
  ``Refraction of geodesics by impulsive spherical gravitational waves in constant-curvature spacetimes with a cosmological constant,''
Phys.\ Rev.\ D {\bf 81}, 124035 (2010).
[arXiv:1005.4613 [gr-qc]].
}

\lref\StephaniTM{
  H.~Stephani, D.~Kramer, M.~A.~H.~MacCallum, C.~Hoenselaers and E.~Herlt,
  ``Exact solutions of Einstein's field equations,''
}

\lref\AdamoVU{
  T.~M.~Adamo, C.~N.~Kozameh and E.~T.~Newman,
  ``Null Geodesic Congruences, Asymptotically Flat Space-Times and Their Physical Interpretation,''
Living Rev.\ Rel.\  {\bf 12}, 6 (2009), [Living Rev.\ Rel.\  {\bf 15}, 1 (2012)].
[arXiv:0906.2155 [gr-qc]].
}

\lref\deFreitasLIA{
  G.~Bernardi de Freitas and H.~S.~Reall,
  ``Algebraically special solutions in AdS/CFT,''
JHEP {\bf 1406}, 148 (2014).
[arXiv:1403.3537 [hep-th]].
}
\lref\PasterskiTVA{
  S.~Pasterski, A.~Strominger and A.~Zhiboedov,
  ``New Gravitational Memories,''
[arXiv:1502.06120 [hep-th]].
}
\lref\PoissonNV{
  E.~Poisson,
  ``A Reformulation of the Barrabes-Israel null shell formalism,''
[gr-qc/0207101].}
\lref\BlauNEE{
  M.~Blau and M.~O'Loughlin,
``Horizon Shells and BMS-like Soldering Transformations,''
JHEP {\bf 1603}, 029 (2016).
[arXiv:1512.02858 [hep-th]].}
\lref\BakasKFA{
  I.~Bakas and K.~Skenderis,
  ``Non-equilibrium dynamics and $AdS_4$ Robinson-Trautman,''
JHEP {\bf 1408}, 056 (2014).
[arXiv:1404.4824 [hep-th]].
}

\lref\BakasHDC{
  I.~Bakas, K.~Skenderis and B.~Withers,
  ``Self-similar equilibration of strongly interacting systems from holography,''
Phys.\ Rev.\ D {\bf 93}, no. 10, 101902 (2016).
[arXiv:1512.09151 [hep-th]].
}
\lref\BarrabesNG{
  C.~Barrabes and W.~Israel,
 ``Thin shells in general relativity and cosmology: The Lightlike limit,''
Phys.\ Rev.\ D {\bf 43}, 1129 (1991).}
\lref\KinnersleyZW{
  W.~Kinnersley and M.~Walker,
  ``Uniformly accelerating charged mass in general relativity,''
Phys.\ Rev.\ D {\bf 2}, 1359 (1970)..
}

\lref\GriffithsTK{
  J.~B.~Griffiths, P.~Krtous and J.~Podolsky,
  ``Interpreting the C-metric,''
Class.\ Quant.\ Grav.\  {\bf 23}, 6745 (2006).
[gr-qc/0609056].
}

\lref\GregoryHD{
  R.~Gregory and M.~Hindmarsh,
  ``Smooth metrics for snapping strings,''
Phys.\ Rev.\ D {\bf 52}, 5598 (1995).
[gr-qc/9506054].
}

\Title{\vbox{\baselineskip12pt}} {\vbox{
\centerline {Superrotations and Black Hole Pair Creation} }} 

\centerline{Andrew Strominger and Alexander Zhiboedov} \vskip.1in \centerline{\it Center for the Fundamental Laws of Nature}\centerline{\it
Harvard University, Cambridge, MA 02138 USA}

\vskip.8in \centerline{\bf Abstract} { 
Recent work has shown that the symmetries of classical gravitational scattering in asymptotically flat spacetimes include, at the linearized level, infinitesimal superrotations. These act like Virasoro generators on the  celestial sphere at null infinity. However, due to the singularities in these generators, the physical status of finite superrotations has remained unclear. Here we address this issue in the context of the breaking of a
cosmic string via quantum black hole pair nucleation. This process is described by a gravitational instanton known as the $C$-metric. After pair production, the black holes are pulled by the string to null infinity with a constant acceleration. At late times the string decays and the spacetime settles into a vacuum state. We show that the early and late spacetimes before and after string decay differ by a finite superrotation. This provides a physical interpretation of superrotations. They act on spacetimes which are asymptotically flat everywhere except at isolated singularities with cosmic string defects. 
}

\Date{}








\newsec{Introduction}

Many years ago, Bondi, van der Burgh, Metzner  and Sachs (BMS) \bms\ looked for diffeomorphisms of an asymptotically flat spacetime which act nontrivially on the physical data of general relativity at null infinity. Rather than the expected Poincare group,  they found  infinite-dimensional asymptotic symmetry groups BMS$-$ (BMS$+$) acting on past (future) null infinity ($\ci^-$(\ip)). These groups are semi-direct products of the Lorentz group and an infinite-dimensional group of `supertranslations' which translate separately along each null generator of $\ci^\pm$. More recently \StromingerJFA\ it was shown that a certain subgroup of the product of the  BMS$+$ and BMS$-$ groups  is a symmetry of gravitational scattering. Moreover this subgroup acts nontrivially on flat Minkowski space, generating an infinite degeneracy of gravitational vacua, all with vanishing ADM mass but differing ADM angular momenta.  The gravitational memory effect can be understood \StromingerPWA\ as physically measuring the transitions between these degenerate vacua. 

Much more recently, it has been conjectured with a variety of motivations \refs{\deBoerVF\BanksVP-\BarnichSE} that BMS missed an infinite number of asymptotic symmetries known as `superrotations'. Roughly speaking, these are symmetries which 
separately rotate around each generator of $\ci$. Superrotations enlarge the $SL(2,C)$ Lorentz group, which acts like the global conformal group on the celestial sphere, to the full Virasoro group . 

A perturbative proof that the {\it infinitesimal} superrotations are a linearized symmetry of classical gravitational scattering was given in  
\refs{\CachazoFWA\KapecOPA\CampigliaYKA\KapecJLD-\CheungIUB}.\foot{As discussed in  \PasterskiTVA\ this is enough to insure superrotation charge conservation, which is measured by the spin memory effect.} However the situation for {\it finite} superrotations has remained unclear, largely because of the singularities at isolated points in the celestial sphere \CompereJWB. 
In this paper we argue that finite superrotations do {\it not} act on the space of asymptotically flat spacetimes. Rather, in the cases we study they map globally asymptotically flat spacetimes to ones which are only locally asymptotically flat \AshtekarAR, with isolated defects on the celestial sphere. The defects are interpreted as cosmic strings piercing null infinity. 
The semiclassical decay and subsequent evolution of cosmic strings via black hole pair nucleation is analytically 
described by $C$-metrics \refs{\gary\GarfinkleEQ\GarfinkleXK\HawkingZN\EmparanJE\EardleyAU-\MaldacenaXJA}.
We show herein that the initial and final metrics in such a process differ by a finite superrotation.  Hence cosmic string decay provides a superrotational analog of the supertranslation vacuum transitions measured by the gravitational memory effect. 

Let us briefly clarify this with a few equations. The large-$r$ expansion near \ip\ of an asymptotically flat metric  in retarded Bondi coordinates takes the form\foot{ $m_B(u,z,\zb)$ is the Bondi mass aspect, $u$ is retarded time, $\gamma_{z\zb}={2 \over (1+z\zb)^2}$ is the metric on the unit sphere and  $D_z$ is the $\gamma$-covariant derivative. 
We follow the conventions of \StromingerPWA.}
\eqn\asymptflat{\eqalign{
d s^2 &= - d u^2 - 2 du dr + 2 r^2  \gamma_{z \bar z} 
dz d \bar z \cr
&+2{m_B \over r} d u^2 + r C_{zz} d z^2 + r C_{\bar z \bar z} d \bar z^2 + D^{z} C_{z z} du dz + D^{\bar z} C_{\bar z \bar z} du d \bar z+...
}}
where the first line is the flat Minkowski metric and the second the leading corrections.
Supertranslations are diffeomorphisms which preserve the asymptotic form of the metric \asymptflat\ while shifting the retarded time 
\eqn\str{u\to u-f(z,\zb),} where $f$ is an arbitrary function on the sphere. Under a supertranslation one finds \eqn\supertranslatedstate{
\delta_f C_{zz} = -  2 D_z^2 f .
}
Two spacetimes which differ by a shift of the form \supertranslatedstate\ are physically inequivalent.  In general, if we disturb a spacetime and let it settle back down, a supertranslation will be induced. Consider a situation where the asymptotic geometry is static before some initial time $u_i$ and after some final time $u_f$, but has nonzero Bondi news $N_{zz}=\p_uC_{zz}$ during the intermediate period
$u_i<u<u_f$. Then, as shown in \StromingerPWA,  the difference $\Delta C_{zz}$ between the  initial and final values of $C_{zz}$ is given by a  supertranslation with\foot{ Here $T_{uu}/r^2$ is energy density in gravitational plus other forms of radiation passing through \ip\ and the Green function is   $G(z,\zb;w,\bar w) =-\left[ {|z-w|^2\over \pi(1 + w \bar w)(1 + z \bar z) } \right] \log\left[ {|z-w|^2\over (1 + w \bar w)(1 + z \bar z) } \right] $.}
\eqn\jump{ f(z,\bar z) = 2 \int_{u_i}^{u_f}du\int d^2 w \gamma_{w \bar w} G(z,\zb;w,\bar w)\left( T_{u u} (u,w,\bar w)  + \p_u m_{B}(u,w,\bar w)  \right) .}
In the shock wave limit  $m_B$ has a discontinuity, the energy flux is a delta function, and the supertranslation occurs instantaneously.

Now we turn to finite superrotations \penrose, \refs{ \deBoerVF\BanksVP-\BarnichSE}. These are diffeomorphisms in which the $S^2$ coordinate transforms as 
\eqn\fed{ z \to w(z)} where $w$ is locally a holomorphic function of $z$. They also preserve the asymptotic form of the metric \asymptflat\ {\it except} at singularities of $w$. 
  In this case (see {\it e.g.} \refs{\penrose, \BarnichLYG,\CompereJWB}) the finite transformation is given in terms of the holomorphic function $w(z)$
\eqn\superrotations{
C_{zz} =- u \{ w , z \}
}
where $\{ w , z \}$ is the usual Schwarzian derivative familiar from 2d CFT \eqn\schwarzian{
\{ w , z \} = {w''' \over w'} - {3 \over 2} \left(  {w'' \over w' }\right)^2 .
}
Following the construction for  supertranslations,  we would like to find  spacetimes in which $C_{zz}$ differs by  \superrotations\ at late and early times.  We demonstrate below, in an semiclassically soluble example,  that a simple example of such a transition  is provided by a cosmic string that snaps in two pieces via black hole pair creation. The finite superrotation is $w=z^{1-4G\mu}$ where $G$ is Newton's constant and $\mu$ is the string tension.  Singularities of the Schwarzian are associated with the endpoints of the string. The fact that the generic holomorphic transformation is not globally well-defined corresponds to the fact that the spacetime changes its topology. Near \ip\  the spacetime is only locally asymptotically flat due to the conical deficit caused by the presence of the cosmic string.

In section 2 we review the Penrose construction \penrose\ of superrotated spacetimes by cutting along the lightcone and regluing with a conformal transformation. In section 3 we discuss how this can be approximately realized \refs{\GleiserVT, \penrosenutku}, up to endpoint singularities by snapping cosmic strings. Section 4 describes the semiclassically soluble and fully consistent example of cosmic string decay via charged black hole pair production \refs{\HawkingZN,\EmparanJE,\EardleyAU}, and show that the decay process mediates between two spacetimes which differ by a finite superrotation. We conclude with a speculative comment on a celestial torus. Appendix $A$ presents the spinning $C$-metric which allows for spin as well as charge on the black holes. Appendix $B$ 
presents the Robinson-Trautman geometries whose form suggests a connection to more general types of superrotations.

\newsec{Penrose Warps and Superrotations}

In this section we review the ``cut and paste'' construction of Penrose \penrose. In this work Penrose constructed local vacuum solutions of the Einstein equation by gluing two patches of Minkowski space along a null surface. In the usual retarded coordinates Minkowski space takes the form
\eqn\minknonus{\eqalign{
d s^2 &=- d u^2  - 2 d u d r  + 2 r^2 \gamma_{z \bar z} d z d \bar z , \cr
\gamma_{z \bar z} &= {2 \over \left( 1 + z \bar z\right)^2 } .
}}
The idea is to cut the space along the  light-cone $u=0$ that is originating  from $r=0$ and then glue it back together after  performing  a diffeomorphism  on the patch with $u>0$ in such a way that the metric is continuous across $u=0$.  The matching condition is simply 
\eqn\matching{
r_-^2  \gamma_{z_- \bar z_-} d z_- d \bar z_- = r_+^2  \gamma_{z_+ \bar z_+} d z_+ d \bar z_+ ,
}
where the subscripts denote upper and lower patches. 
This is locally solved by 
\eqn\solutionwrap{\eqalign{
z_+ &= w(z_-) , \cr
r_+ &= {1 + w(z_-) \bar f(\bar z_-) \over 1 + z \bar z } {1 \over \left( w'(z_-) \bar w'(\bar z_-) \right)^{{1 \over 2}} } r_-  \ ,
}}
where $w(z)$ is an arbitrary holomorphic function. These are conformal transformations: the change in the sphere metric  $r^2\gamma_{z\zb}$ induced by $z\to w(z)$ is cancelled by an angle-dependent shift in the radial coordinate. 

It is possible to introduce coordinates that are continuous across $u=0$. One finds  \refs{\penrosenutku,\HoganXJ}
\eqn\fullmetric{\eqalign{
d s^2 &= - 2 d u d r - d u^2 + \left( 2 r^2 \gamma_{z \bar z} + {1 \over 4} u^2 \theta(-u) (1+ z \bar z)^2 \{ w , z \}  \{\bar  w , \bar z \} \right)d z d \bar z \cr
&- r u \theta(-u) \left(  \{ w , z \} d z^2 + \{ \bar w , \bar z \} d \bar z^2 \right) , \cr
}}
where $\{ w(z), z \}$ is again the  Schwarzian and $\theta(-u)$ vanishes for positive $u$ and is otherwise equal to one. By construction this metric is flat for $u\neq 0$. It turns out that the Ricci tensor is zero everywhere, including $u=0$, except at $r=0$ and the singularities of $w(z)$ which we discuss below.\foot{To avoid possible confusion let us mention that \fullmetric\ is not written in the Bondi gauge. It can be, however, transformed to the Bondi gauge with $C_{zz} =- u \theta(-u) \{ w , z \}$ as above. See also appendix B.} 


We have also have a delta function in the Weyl tensor for $u=0$ which takes the form
\eqn\riemann{
C_{u z u z} =-  {r \over 2} \delta(u) \{ w , z \}  .
}
These are a form of the gravitational impulse waves studied in, for example, \refs{\penrose, \BarrabesNG\PodolskyNN \PoissonNV-\BlauNEE}.

\newsec{Snapping Cosmic String}

In his original work, Penrose \penrose\ noticed that globally the transformation $z \to w(z)$ introduces singularities which he called ``wire singularities.'' Later \refs{\GleiserVT, \penrosenutku} it was shown that for  simple choice of $w(z) = z^{\alpha}$ the solution above describes the gravitational shock wave produced by the snapping of  a cosmic string in which the two ends start at $r=u=0$ and  travel along the singularities at $z=(0,\infty)$ at the speed of light out to null infinity.
More precisely, the metric around a cosmic string in the zero width approximation is \VilenkinZS
\eqn\cosmicstring{
d s^2 = - d u'^2  - 2 d u' d r'  + r'^2 (d \theta'^2 + (1 - 4 G \mu)^2 \sin^2 \theta' \ d \phi^2 )\ , ~~~ 0 \leq \phi < 2 \pi  .
}
where $\mu$ is the energy density per unit length of the string and $G$ is Newton's constant. The geometry is everywhere flat except along the location of the string on the polar axis $\theta = 0, \pi$ where there is a deficit angle deficit angle \eqn\hty{\delta=8 \pi G \mu.} 
This geometry is not asymptotically flat because the strings pierce null infinity. They are however asymptotically locally flat (ALF) in the sense of \AshtekarAR. The unexcited cosmic string \cosmicstring\ can be viewed as an ALF vacuum with different asymptotics than the stringless vacuum.   We could for example study scattering in the presence of such  defects. 

Bondi coordinates $(u,r ,\theta,\phi)$ are given in a  large-$r$ expansion by the expression \refs{\BicakRB,\Bicak}\eqn\changetobondi{\eqalign{
u'  &=  {u \over W} + ..., \cr
r' &= r W- u {1 - W^2 - (\pa_\theta W)^2 \over 2 W} + ... , \cr
\theta' &= \int^\theta {d x \over W(x)}- {u \over r} \pa_{\theta} {1 \over W}+ ... ,\cr 
W&\equiv {1 \over (1 - 4 G \mu)} {  \sin \theta \over \sin \theta'}  \ .
}}
The leading order relation between $\theta$ and $\theta'$ is 
\eqn\relation{
\pa_\theta \theta' = {1 \over W} \ .
}
After this  change of variables,  a subleading  correction to the metric on the sphere appears in the expansion \asymptflat\
\eqn\correction{
- 8 G \mu (1 - 2 G \mu) u  r { d \theta^2 - \sin^2 \theta d \phi^2 \over \sin^2 \theta } = - 4 G \mu (1 - 2 G \mu) u  r \left( {d z^2 \over z^2} + {d \bar z^2 \over \bar z^2} \right)
}
where $z=e^{i\phi}\tan {\theta \over 2}$. 
Equivalently
\eqn\newsofstring{
C_{zz} =- {4 G u \mu (1 - 2 G \mu) \over z^2} = - {u \over z^2} {\delta \over 2 \pi} \left(  1 - {\delta \over 4 \pi} \right).
}
This implies that the Bondi news and mass aspect\foot{To compute the mass aspect one should include further terms in the ${1 \over r}$ expansion in \changetobondi .}
are nonzero everywhere on \ip:
\eqn\newsfstring{\eqalign{
N_{zz} &= -{4 G \mu (1 - 2 G \mu) \over z^2} , \cr
m_B &=-{1 \over 4} u N_{zz} N^{zz} .
}}

Recently the metric with sources generated by finite superrotations was found in  \CompereJWB. The metric of a cosmic string in the Bondi gauge corresponds to $C=0$ and $G(z) = z^{1-4 G \mu}$ in formula $(22)$ of  \CompereJWB .  However such solutions were discarded  in \CompereJWB\ and here we identify them with cosmic strings.

Now consider the superrotation characterized by 
\eqn\simpletrans{
w(z) = z^{1 - 4 G \mu} ,}
Using \superrotations\  one finds that this maps flat Minkowski space to
\eqn\newsoring{
C_{zz} = -u\{w,z\}=-{4 G u \mu (1 - 2 G \mu) \over z^2} .}
Comparing to  \newsofstring\  reveals  that a cosmic string can be understood as the superrotation of flat space given by 
\simpletrans. Superrotations map one flat  geometry to another flat geometry everywhere except at the singularities. These singularities are cosmic strings which destroy asymptotic flatness except at special points.

In Bondi coordinates the end points of the string are located at $z=0, \infty$.  At $u=0$ the string snaps with the free ends moving with the speed of light to null infinity and for $u>0$ we have Minkowski space with $C_{zz} = 0$.\foot{This decay is not necessary an instantaneous event and can have arbitrary profile in $u$, see e.g. \GriffithsHJ .}

The effect of such gravitational shock waves on test bodies was studied, for example, in \PodolskyXH . In particular, there is a non-zero velocity kick between neighboring geodesics $\Delta v  \sim \Delta N_{zz} \delta z^2$ after passage of the shock wave which signalizes non-zero superrotation and is absent otherwise, see \StromingerPWA\ for details.

Different locally holomorphic functions describe different string configurations. For example, a string along the $x$ axis is described by
\eqn\xstring{
w(z) = \left({1 + z \over 1 - z} \right)^{1- 4 G \mu} .
}

One may also easily construct multi-string configurations as well as collision of cosmic strings \PodolskyCX. We note that for an orbifold space of integer order $n$ such as the `stringy cosmic strings' of \refs{\GreeneYA,\PolchinskiIA} one has 
\eqn\fre{\delta=2\pi \left(1-{1 \over n} \right),~~~~z=w^n.}

Since a cosmic string is a superrotation of flat space, a process in which a cosmic string disappears at some finite retarded time and decays to the stringless vacuum is a domain wall separating two ALF vacua. Transtions between different ALF vacua are characterized by superrotations in the same manner that transitions between BMS vacua are characterized by supertranslations.

\newsec{Charged Black Hole Pair Creation}

The preceding section described the geometry of  a cosmic string that snaps, creating two endpoints that travel to \ip\ at lightspeed. It was not however, an exact solution of the Einstein equation. Solving the Einstein equation exactly will typically lead to gravitational radiation and a change in the Bondi mass. The possibility of such extra effects muddies the relation between cosmic string snapping and superrotations.

Fortunately for us, families of exact classical solutions which describe pairs of various types of black holes leashed at the ends of various types of broken strings and being accelerated to infinity have been known for some time \gary.  These are examples of  lorentzian $C$-metrics \KinnersleyZW\ and have been extensively studied for a variety of reasons \PravdaVH. Moreover quantum tunneling events in which a pair of such leashed black holes  is created is semiclassically described by a euclidean $C$-metric \refs{\gary\GarfinkleEQ\GarfinkleXK\HawkingZN\EmparanJE\EardleyAU-\MaldacenaXJA}. 
The process resembles Schwinger pair production: energy is conserved because the energy of the black hole pair equals the energy of the missing post-nucleation string segment. 

We are particularly interested here in the specific example studied in \refs{\HawkingZN\EmparanJE-\EardleyAU}
describing the nucleation of a pair of oppositely-charged\foot{We will see below that the charge is required for thermal equlibrium.}  black holes leashed on the end of a 
zero-width cosmic string described by a deficit angle. 
The electrically charged C-metric that describes this process is given by \HongGX\foot{We note that  this considerably simplified form of the metric was found in  \HongGX\ subsequent to 
the analyses in \refs{\HawkingZN\EmparanJE-\EardleyAU} which employed more complicated expressions. One can also consider the smoothened version of this process \GregoryHD.}
\eqn\cmetric{\eqalign{
d s^2 &= {1 \over A^{2} (x-y)^{-2}} \left( G(y) dt^2  -  {d y^2\over G(y)}  +{d x^2\over G(x)} +G(x)\kappa^2 {d \phi^2}\right) \ , \cr
G(\zeta) &=(1 - \zeta^2)(1+r_+A\zeta)(1+r_-A\zeta ), \cr~~~{\cal A}&= q y dt, ~~~{\cal F}=d{\cal A} }}
where \eqn\fixingkappa{
\kappa = {2 \over |G'(\zeta_4)|}  , 
}
\eqn\sw{ r_\pm=m\pm \sqrt{m^2-q^2},~~~0 <r_- A<r_+A<1.}
The four real roots of $G$
are
\eqn\eh{\zeta_1=-{1\over r_-A},~~\zeta_2=-{1\over r_+A},~~\zeta_3=-1,~~\zeta_4=1.}
This solution describes  a patch of a  spacetime in which two black holes of masses $m$ and charges $\pm q$ are dragged to null infinity by a pair of cosmic strings. The `Rindler wedge' associated with one of the black holes is the region
\eqn\eqa{ \zeta_2\le y \le \zeta_3,~~~\zeta_3\le x \le \zeta_4.}
The event horizon is $y=\zeta_2$ and the acceleration horizon is $y=\zeta_3$. The region past the acceleration horizon is described by $\zeta_3 \le y \le x$.\foot{The global structure of the solution is reviewed, for instance, in \GriffithsTK .}
In order to avoid a conical singularity at $x=\zeta_4$ between the black holes, $\phi$ is identified, with our defintion of $\kappa$,  as
\eqn\wwt{\phi \sim \phi+2\pi.}
This leads to a deficit angle 
\eqn\yop{\delta=2\pi\left(1- \left|{G'(\zeta_3)\over G'(\zeta_4)}\right| \right)}
along $x=\zeta_3$ corresponding to a cosmic string stretching to infinity. 

This solution has 3 parameters $(A,m,q)$: the acceleration, mass and charge of the black hole. Generically the Hawking and Unruh temperatures of the black hole will not be equal and the solution will be quantum mechanically unstable. Equating these is equivalent to demanding the equality of the surface gravities associated to the Killing vector $\p_t$ at the two horizons, or $G'(\zeta_2)=-G'(\zeta_3)$.\foot{This is equivalent to demanding smoothness of the Euclidean section \GarfinkleEQ.} This is solved
 by 
 \eqn\ijj{A={r_+-r_- \over 2r_+r_-}={\sqrt{m^2-q^2}\over q^2}.}
 
 This leaves a two parameter family of solutions labelled by say the deficit angle of the cosmic string and the mass $m$ of the black hole.  These solution are all energetically degenerate because the energy of the black hole pair equals 
 the energy of the `missing" segment of cosmic string. Using \ijj\  one finds 
 \eqn\fixingkappa{
\kappa =  {q^2 \over m^2 + 2 m \sqrt{m^2 - q^2}} , 
}
 and the deficit angle becomes
 \eqn\ser{\delta ={8\pi m A\over 1+ 2mA +q^2A^2} = {8 \pi \sqrt{m^2 - q^2} \over m+ 2 \sqrt{m^2 - q^2}}.}
 We note that for $m\to q$, the black hole becomes near-extremal and the deficit angle and
 acceleration become small.

The solution possesses rotation and boost symmetry. When written in the Bondi gauge these symmetries constrain the possible form of the news tensor to be \BicakVZ
\eqn\symmetry{\eqalign{
N_{\theta \theta} &= - {N_{\phi \phi} \over (\sin \theta)^2 } = {f({u \over \sin \theta}) \over (\sin \theta)^2}  , \cr
{ N_{\theta \phi} \over \sin \theta} &= {g({u \over \sin \theta}) \over (\sin \theta)^2} ,
}}
where $f$ and $g$ take particular values for a given solution. For the charged solution above $N_{\theta \phi} = 0$ (but not for its spinning generalization). Switching to $z, \bar z$ coordinates gives 
\eqn\genericnews{
N_{zz} = {f -ig\over z^2} ,
}
where ${u \over \sin \theta}={u (1 + z \bar z) \over 2 \sqrt{ z \bar z} }$. One can check that \genericnews\ is invariant under the rotation and the boost Killing symmetries
\eqn\symmetries{\eqalign{
\zeta^{\mu} \pa_{\mu} &= Y^z \pa_z + {u \over 2} D_z Y^z \pa_u + c. c. \  , \cr
Y^{z} &= i z, ~~~ Y^z = z \ .
}}

Notice that \symmetry\ implies that asymptotics $u \to \pm \infty$, $\theta$ fixed, and $\theta \to 0, \pi$, $u$ fixed are controlled by the same function $f(x)$. Thus, if we know the behavior of the news in the far past and the far future we can also understand it close to $\theta = 0, \pi$ and vice versa. In our case $\theta = 0, \pi$ correspond to the locations of the cosmic string.

To compute the precise form of the functions $f$ and $g$ in the case of the metric above we can use the elegant method of \tafel\  applied to the C-metric with zero charge in \SladekWY. The only change in their analysis for the case of the charged metric is the form of $G(x)$ function. Thus, we can simply quote the result of \SladekWY\ with $G(x)$ defined above \cmetric
\eqn\result{
N_{\theta \theta} = - {1 \over (\sin \theta)^2} (1 + {\kappa^2 \over 2} G G'' - {\kappa^2 \over 4} (G')^2 ) .
} 
Here $G(x)$ is a function of ${u \over \sin \theta}$ as argued above from symmetries considerations  \symmetry . The relation between $x$ and ${u \over \sin \theta}$ is given by 
\eqn\relation{
{u \over \sin \theta} = {1 \over A \kappa} \int_{0}^x {d x' \over G(x')^{3/2}} \ .
}
In writing this we fixed the supertranslation frame such that the black holes arrive to ${\cal I}^+$ at $u=0$.

It is very easy to extract some qualitative features of the news tensor. Recall that for $u<0$ we expect to have singularities due to the presence of the string and for $u>0$ we expect the news to be regular. 
Consider first the case $u<0$ in this case $\theta = 0, \pi$ correspond to $x=\zeta_3$ in \relation. Plugging this in \result\ we get for the leading asymptotic
\eqn\leading{
N_{\theta \theta} = - {8 m \sqrt{m^2 - q^2} \over (m+ 2 \sqrt{m^2 -q^2})^2} {1 \over (\sin \theta)^2} + ... \ . 
}
Based on previous discussion in the presence of the cosmic string we expect the result to be \correction\ $- {\delta \over \pi} (1 - {\delta \over 4 \pi}) {1 \over (\sin \theta)^2}$. Plugging \ser\ for $\delta$ we indeed reproduce \leading .
For $u>0$ $\theta = 0, \pi$ corresponds to $x=\zeta_4$ in \relation. Plugging this in \result\ we get $0$ to leading order as expected. The first non-zero term behaves as ${(\sin \theta)^2 \over u^4}$. 

Similarly, by analyzing the same limit we can estimate the behavior of the news tensor for fixed $\theta$ and $u \to \pm \infty$. By the same argument we get that for $u \to - \infty$ it is given by the news tensor of the cosmic string, whereas for $u \to \infty$ it goes to $0$ 
\eqn\limitsnews{\eqalign{
\lim_{u \to - \infty}N_{\theta \theta}&= - {8 m \sqrt{m^2 - q^2} \over (m+2  \sqrt{m^2 -q^2})^2} {1 \over (\sin \theta)^2} , \cr
\lim_{u \to \infty} N_{\theta \theta} &= 0 .
}}
These asymptotics are related by a finite superrotation as described in the previous sections. 

\newsec{Concluding Comment}
We conclude with a speculative comment. 

Multi-particle scattering amplitudes of any 4D quantum field theory may always be recast as two-dimensional operator correlation functions on the celestial sphere at null infinity \HeZEA. The operators are labelled by the location at which the particle pierces the celestial sphere as well as other quantum numbers such as the energy.\foot{In the massive case the location is  integrated  over a region on the sphere \CampigliaQKA, \KapecENA.}  The 4D $SL(2,C)$ Lorentz symmetry implies invariance of these correlators under the global conformal symmetry of the sphere. 
In 4D quantum gravity, the global conformal symmetry is enhanced at tree-level to the full local one \refs{\CachazoFWA\KapecOPA\CampigliaYKA\KapecJLD-\CheungIUB}.  
This implies 4D quantum gravity scattering amplitudes behave like those of a CFT$_2$, 
although presumably they are in very different representations than the familiar unitary varieties. We can imagine that there is  a holographically dual representation of this CFT$_2$ on the celestial sphere which is intrinsically two-dimensional.

 It is natural to ask then what happens to the celestial sphere when it is pierced by black holes accelerated out to  $\ci^+$ by cosmic strings. We suggest here  that it turns into a celestial  torus. 
Consider the scattering process in which an ingoing cosmic string decays into a pair of black holes which reach the celestial sphere \ip, one at the north pole and one at the south pole.   The outgoing state is an eigenstate of asymptotic boosts $L_0+\bar L_0$ and rotations $L_0-\bar L_0$ which leave fixed the two points at which the black holes emerge.  Geometrically, the two black holes are connected by a wormhole which turns the sphere on which they emerge into a torus.  In the dual CFT$_2$, 
each black hole corresponds to an operator insertion on the sphere at the point where it emerges, with the operator depending on the particular microstate the black hole is in.  As shown  in 
\GarfinkleXK\ and employed recently in \MaldacenaXJA , the individual black holes are in thermal ensembles 
at the Hawking-Unruh temperature, but the pair of black holes are correlated so as to form a pure state. This was deduced in \GarfinkleXK\ from identifying the black hole entropy contribution to the decay rate. In the dual CFT$_2$ picture, such a scattering amplitude is represented by a weighted insertion of pairs of all possible operators with one at the north and the other at the south pole. 
The weighting factor includes\foot{Charged fields would have an additional weighting factor proportional to the electrostatic potential at the horizon.}  \eqn\wft{e^{-\beta(L_0+\bar L_0) },} where $L_0\pm\bar L_0 $ are the boost and rotation eigenvalues of the operators and $\beta$ is the inverse of the Hawking-Unruh temperature. In a well-behaved CFT$_2$, such a weighted insertion of operators turns the sphere into a torus with modular parameter $i\beta$. In this picture the cosmic string decay rate is partially computed by a partition function in the dual CFT$_2$ which should contain a factor of the black hole entropy. It would be fascinating indeed if this could be made precise! 
\newsec{Acknowledgements}
We are grateful to Gary Horowitz and Malcolm Perry for useful discussions. This work was supported by NSF grant 1205550.

\appendix{A}{Spinning Black Hole Pair Creation}
We record here the spinning  C-metric to which a similar analysis applies with charged black holes replaced by rotating ones. The metric  is
\eqn\cmetric{\eqalign{
d s^2 &= {1 \over A^{2} (x-y)^{-2}} \bigl( {G(y)\over 1+(aAxy)^2} (dt-aAx^2d\phi)^2  -  {1+(aAxy)^2\over G(y)}  d y^2\cr&~~~~~~~~~~~~~~~~~+{ 1+(aAxy)^2\over G(x)}d x^2 +{G(x)\over  1+(aAxy)^2}\kappa^2 (d \phi+aAy^2dt)^2\bigr) \ , \cr
G(\zeta) &=(1 - \zeta^2)(1+r_+A\zeta)(1+r_-A\zeta ),   }}
where \eqn\fixingkappa{
\kappa = {2 \over |G'(\zeta_4)|}  , 
}
\eqn\sw{ r_\pm=m\pm \sqrt{m^2-a^2},~~~0 <r_- A<r_+A<1.}
The four real roots of $G$
are
\eqn\eh{\zeta_1=-{1\over r_-A},~~\zeta_2=-{1\over r_+A},~~\zeta_3=-1,~~\zeta_4=1.}
This solution describes  a patch of a  spacetime in which two black holes of masses $m$ and charges $\pm q$ are dragged to null infinity by a pair of cosmic strings. The `Rindler wedge' associated with one of the black holes is the region
\eqn\eqa{ \zeta_2\le y \le \zeta_3,~~~\zeta_3\le x \le \zeta_4.}
The ergosphere begins at $y=\zeta_2$ and the acceleration horizon is $y=\zeta_3$. The region past the acceleration horizon is described by $\zeta_3 \le y \le x$.
In order to consider the conical singularities, it is useful to define
\eqn\dsi{\tilde t=t-aA\phi,} in terms of which the metric becomes 
\eqn\cmetric{\eqalign{
d s^2 &= {1 \over A^{2} (x-y)^{-2}} \bigl( {G(y)\over 1+(aAxy)^2} (d\tilde t+aA(1-x^2)d\phi)^2  -  {1+(aAxy)^2\over G(y)}  d y^2\cr&~~~~~~~~~~~~~~~~~+{ 1+(aAxy)^2\over G(x)}d x^2 +{G(x)\over  1+(aAxy)^2}\kappa^2 ((1+a^2A^2y^2)d \phi+aAy^2d\tilde t)^2\bigr) \ ,   }}
It is then evident that at $x=\zeta_4=1$ between the black holes, there is no singularity if $\phi$ is identified $\phi \sim \phi+2\pi$ (with $\tilde t$ held fixed) while at $x=\zeta_3=-1$ there is 
a deficit angle \yop.        

As in the charged black hole analysis, the superrotation associated to spinning black hole pair creation depends only on the initial cosmic string deficit angle and is independent of the detailed process by which it disappears.
 
\appendix{B}{Robinson-Trautman Spacetimes}

The $C$-metrics described above are particular cases of the Robinson-Trautman (RT) solution \RobinsonZZ, which may have applications to the study of superrotations and are recorded in this appendix.\foot{This does not include the case of the spinning C-metric. } The RT solution describes a space that admits a twist and shear-free null geodesic congruence (for review see \StephaniTM). One can think of these spacetimes as describing radiation from the bounded sources \ChruscielCJ . 

The RT metric takes the form
\eqn\robtraut{
d s^2 = -h(u,z, \bar z) du^2  - 2 d u d r + 2 {2 r^2\over P(u, z, \bar z)^2} d z d \bar z ,
} 
where
\eqn\robtrauttwo{
h(u,z , \bar z) = P^2 \pa_z \pa_{\bar z} \log P - 2 r \pa_u \log P - {2 m(u) \over r}  \ 
}
and the equations of motion reduce to
\eqn\eomrobtraut{
- 4 \pa_u m + 12 m \pa_u \log P + P^2 \pa_z \pa_{\bar z} \left( P^2 \pa_z \pa_{\bar z} \log P \right) =16 \pi G r^2 T_{u u} . 
}

The news tensor for this class of solutions was computed in \refs{\vonderGonnaSH\Cornish\tafel-\AdamoVU} with a simple result
\eqn\simpleresult{\eqalign{
N_{zz}(u_B, z, \bar z) &= 2 {\pa_z^2 P(u,z,\bar z) \over P(u, z, \bar z)} |_{u = u (u_B, z , \bar z)}\ , \cr
u_B &= \int_{0}^{u} {P(U, z, \bar z) \over 1 + z \bar z} d U \ ,
}}
where $u_B$ is the Bondi retarded time and taking $z$ derivatives in the first line should be done first and then $u$ should be substitutes in terms of $u_B$ and $(z, \bar z)$. In the second line we chose a particular supertranslation frame.

A simple solution that generalizes the spacetimes that we discussed in the previous section corresponds to $m = T_{uu} = 0$ and $P(u, z , \bar z) = { 1 + f(u,z) \bar f(u, \bar z) \over (\pa_z  f \pa_{\bar z} \bar f)^{1/2}}$.\foot{More generally, we can consider non-zero $m(u)$ and $T_{uu}$ that would satisfy \eomrobtraut.} This is the most general type N Robinson-Trautman solution. The news tensor in this case takes the form 
\eqn\newstensor{\eqalign{
N_{zz}(u_B, z, \bar z) &=- \left\{ f(u,z) , z \right\} |_{u = u (u_B, z , \bar z)} \ .
}}
For $\pa_u f(u,z) = 0$ the spacetime is locally flat. Robinson-Trautman solutions in higher dimensions were studied in \PodolskyDU  . In the context of AdS/CFT these solutions were recently studied in \refs{\deFreitasLIA\BakasKFA-\BakasHDC}. It would be interesting to understand if there is any connection between the physics at null infinity and AdS boundary in this case.
 
\listrefs

\bye